# A concise review on THGEM detectors

A.Breskin [a,*], R. Alon [a], M. Cortesi [a], R. Chechik [a], J. Miyamoto [a]

V. Dangendorf [b], J. Maia [c,d] and J. M. F. Dos Santos [c]

[a] *Dept. of Particle Physics, The Weizmann institute of science, 76100 Rehovot, Israel*

[b] *Physikalisch-Technische Bundesanstalt (PTB), Braunschweig, Germany*

[c] *University of Coimbra, Portugal*

[d] *University of Beira Interior, 6201-001 Covilhã, Portugal*

**Abstract**

We briefly review the concept and properties of the Thick GEM (THGEM); it is a robust, high-gain gaseous electron multiplier, manufactured economically by standard printed-circuit drilling and etching technology. Its operation and structure resemble that of GEMs but with 5 to 20-fold expanded dimensions. The millimeter-scale hole-size results in good electron transport and in large avalanche-multiplication factors, e.g. reaching $10^7$ in double-THGEM cascaded single-photoelectron detectors. The multiplier's material, parameters and shape can be application-tailored; it can operate practically in any counting gas, including noble gases, over a pressure range spanning from 1 mbar to several bars; its operation at cryogenic (LAr) conditions was recently demonstrated. The high gain, sub-millimeter spatial resolution, high counting-rate capability, good timing properties and the possibility of industrial production capability of large-area robust detectors, pave ways towards a broad spectrum of potential applications; some are discussed here in brief.

[*] Corresponding author. Tel.: 00972-8-9342645; fax: 00972-8-9342611; e-mail: amos.breskin@weizmann.ac.il.

## 1. Introduction

Gaseous avalanche radiation-imaging detectors have seen subject to intensive developments over the past decades. The so-called micropattern detectors, produced by different micro-lithographic techniques provide localization resolutions in the few-tens of micrometers range, approaching that of silicon trackers [1]. The most advanced operative micropattern detectors are cascaded Gaseous Electron Multipliers (GEM) [2, 3] and the Micromegas [4, 5]. Within the broad family of micropattern gas detectors, the THick Gaseous Electron Multiplier (THGEM) is one of the most recent developments [6]; it has been attracting significant attention due to its simplicity and robustness. The THGEM has a hole-structure similar to the GEM, but with about 10-fold expanded dimensions (Fig.1). It is manufactured economically by mechanically drilling sub-millimeter diameter ($d$) holes, spaced by a fraction of a mm ($a$) in a thin ($t$, generally a fraction of a mm) printed-circuit board (PCB), followed by Cu-etching of the hole's rim (typically 0.1mm). In addition to the standard etching using photolithographic masks (e.g. our THGEMs were manufactured by this process by Print Electronics Inc., Israel. www.print-e.co.il ), a simpler mask-less etching technique was recently proposed and is under investigations [7]. The etched rim reduces edge discharges; this resulted in about ten-fold higher gains than without a rim (Fig.2), e.g. like in the "optimized GEM" [8] or the "LEM" [9].

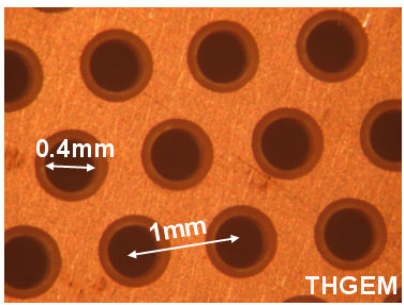

*Figure 1. Photograph of a typical THGEM electrode; the one shown has a hole-diameter of d=0.4mm with 0.1mm etched rim, spaced by a=1mm. The thickness is t=0.5mm.*

Two or more THGEM elements can be cascaded, to provide higher gains or increase operation stability. THGEMs may be fabricated out of various PCB materials - e.g. FR-4, Kevlar, Cirlex (polyimide with low natural radioactivity [10]), Teflon etc. Due to their mechanical robustness, THGEM-based detectors may be constructed with very large area and their implementation does not require any particular mechanical supports.

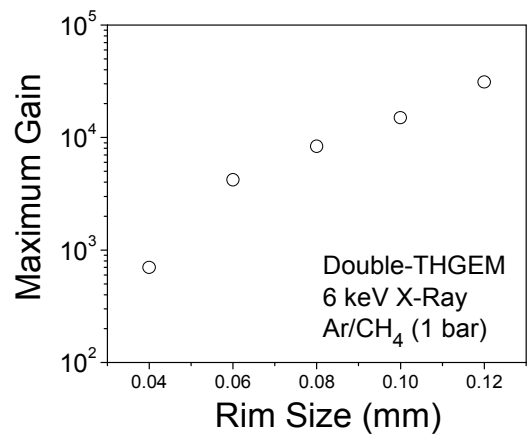

*Figure 2. Maximum attainable gain versus rim size. Detector parameters: t=0.4 mm; a=1 mm; d=0.3 mm.*

In this work we briefly review the operation-principle of THGEM detectors and their properties. Details can be found in earlier articles [11-15] and theses [16,17]. Some recent results on time resolution and operation in noble gases as well as potential applications are briefly discussed.

## 2. THGEM operation and properties

*General*

The THGEM's operation principle is basically the same as that of the GEM: an electric potential is applied between the electrodes and creates a strong dipole electric field within the holes, protruding also into the adjacent volume. This particular shape of the field is responsible for an efficient focusing of ionization electrons into the holes and their multiplication by a gas avalanche process. The electron collection is more effective than in GEM because the THGEM's hole-diameter is larger than the electron's transverse diffusion range when approaching the hole. The efficient collection and transmission of electrons offers the possibility to use several THGEM elements in cascade. This leads to higher detector gains at lower voltage bias per single THGEM element and thus to higher operation stability. This is important when the detected radiation has a large dynamic range in primary ionisation density (e.g. neutrons, radioactive background etc.). The THGEM can efficiently detect radiation-induced electrons, either deposited in the gas or emitted from a solid converter. While the former is important for particle tracking, x-ray imaging (Fig. 3) etc, the latter has important applications in single-photon [13] and neutron imaging.

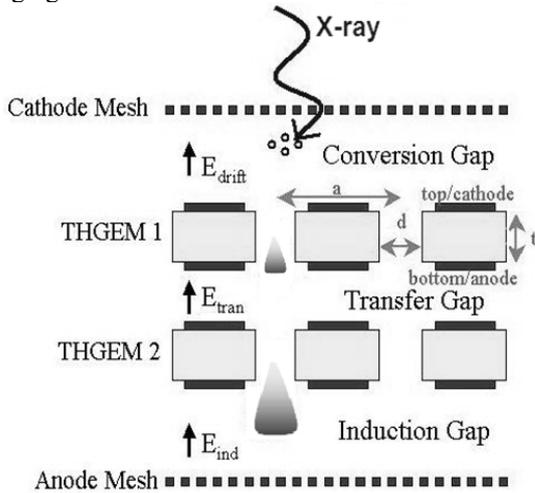

*Figure 3. Schematic view of a double-THGEM soft x-ray detector; the same configuration is adequate for particle tracking and timing.*

The results of systematic studies of THGEM-based detectors, operating at atmospheric and low gas pressures, have been extensively reported in [11,12]. The role of various geometrical and operational parameters, optimal conditions for reaching full single-photoelectron detection efficiency and maximal electron transport were established. The last two prerequisites are particularly important for applications necessitating efficient photon-counting and -imaging with solid photocathodes, as in Cherenkov Ring Imaging detectors (RICH). It was found that due to the large hole's size, efficient electron transport and negligible photon- and ion-feedback, the THGEM has stable operation in a large variety of gas mixtures, including noble gases. High attainable gains, $>10^4$ and $>10^6$, were reached with single photoelectrons in single- and in double-THGEM detectors, respectively, at 1atm of $Ar/5\%CH_4$ and $Ar/30\%CO_2$, thus assuring good sensitivity for single-photon detection. [11,13]. The same detectors yielded gains of $>10^3$ and $>10^4$ in single- and double-THGEM arrangements, respectively, with few-hundred primary electrons induced by 5.9 keV x-rays in 1atm $Ar/5\%CH_4$ [14] (Fig. 4a). In this gas the THGEM reached counting-rate capabilities $>1MHz/mm^2$ at effective gains of $\sim10^4$ [11].

The localization resolution was studied with a 10x10 $cm^2$ double-THGEM detector irradiated with 8 keV x-rays. It comprised two THGEM electrodes of $t$=0.4mm, $d$=0.5mm and $a$=1mm, coupled to a resistive anode; the latter broadened the induced signals, to match the 2mm pitch of the X-Y delay-line readout electrode placed behind it. Localization resolutions of ~0.7 FWHM (smaller than the hole-pitch) were reached in 1bar $Ar/5\%CH_4$ at a gain of $10^4$; the gain variation was of 10% FWHM over the whole surface [14].

*Noble gases and low temperatures*

Gains above $10^4$ were recently measured at room-temperature in a double-THGEM with 5.9 keV x-rays in 1 bar Ar, Kr, Xe, Ne and Ar/5%Xe (Fig. 4b) [18]; gains $>10^3$ were also reached in some of these gases at 2-3 fold higher pressures [18,19]. The energy resolution dependence on various parameters (gas type, pressure, electrode's geometry and electric fields) was studied in detail in noble gases, yielding in some configurations values below 20% FWHM for 5.9 keV x-rays [18].

Recent studies indicated that double-THGEM detectors operated in two-phase liquid Ar could reach gains of ~$10^4$ [20]. The successful operation of THGEM detectors in cryogenic conditions was also reported in [21, 22]. Slower signal development compared to that in cascaded-GEM multipliers was observed in the two-phase operation mode; it permitted noise reduction by pulse-shape analysis and thus lower detection thresholds [20].

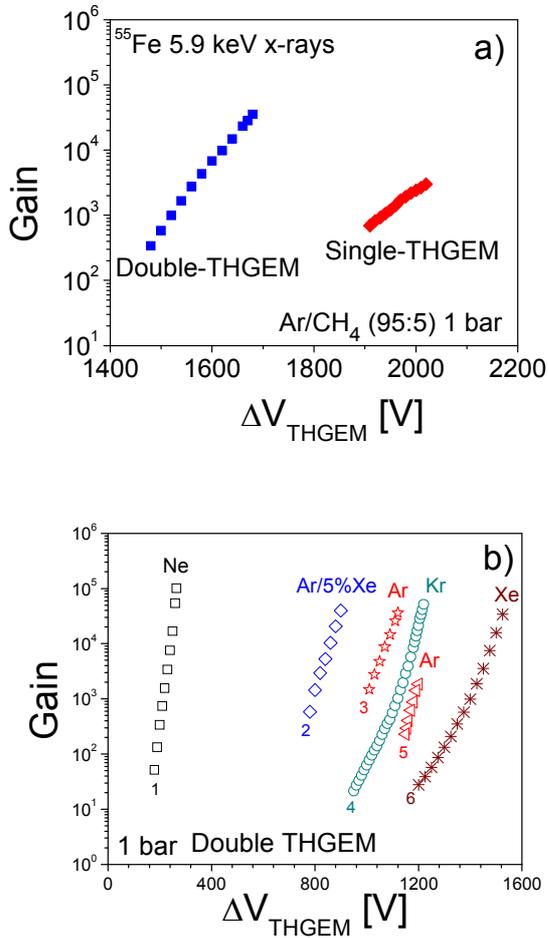

*Fig. 4. Gain curves with 5.9 keV x-rays. a) in single- and double-THGEM (t=0.8mm, d=0.6mm, a=1mm) in 1 bar Ar/5%CH$_4$ ; b) in double-THGEM in noble gases at 1 bar. In a), except for Ar curve (3), measured in gas-flow mode (not purified), all other data were measured with getter-purified gases [18].*

*Rim effects and stability*

As discussed above and shown in Fig. 2, the size of the etched rim around the THGEM holes, is essential for reducing significantly discharge-occurrence probability; this permitted operation at higher permissible voltages and hence at higher detector gains. The relationship between maximum gain versus rim-size was investigated with a 3x3 cm$^2$ double-THGEM, made from standard FR4 printed-circuit board material, operating in Ar/CH$_4$ at atmospheric pressure (setup shown in Fig. 3). It was irradiated with a collimated (1 mm$^2$) $^{55}$Fe x-ray beam. The maximum attainable gain was defined as the one at which micro-discharges were not observed for at least 20 seconds. The maximum attainable gain increased practically exponentially with the rim-size. This effect, as well as the gain stability in time, is due to a combination of several factors: electric field distribution outside the hole, charging up of the insulator, type of material, quality of hole's wall-surface, the surface-quality of the Cu-edge etc. Preliminary results indicated that gain-stabilization with time occurs within a few hours [15]. The matter is being thoroughly investigated in cooperation with CERN and INFN-Trieste within the CERN-RD51 collaboration.

*Time resolution*

The time resolution of a double-THGEM operating in 1bar Ar/5%CH$_4$ at room temperature was measured with UV photons (pulsed UV lamp) and with minimum ionizing charged particles (MIPs).

The detector assembled for UV-photon studies (Fig. 5) had a CsI photocathode, either evaporated on a transparent quartz plate installed at a 3 mm distance from the first THGEM (semi-transparent photocathode), or evaporated directly onto the top surface of the first THGEM (reflective photocathode). Absorbers were used to adjust the number of incident photons per pulse. Both configurations yielded rather similar gain and time-resolution results [17].

Fig. 6 depicts the time resolution measured with a reflective photocathode (Fig. 5). It varied between 8ns and 1ns (RMS) for 1 to ~100 photoelectrons per

UV-pulse. The time resolution for 1000-photoelectron pulses was about 0.5ns (RMS) [17].

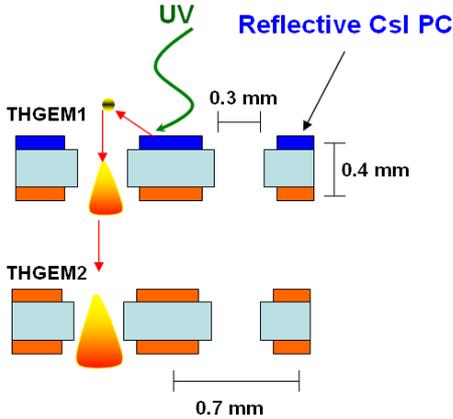

Figure 5. Schematic view of a double-THGEM with a reflective CsI PC deposited on the top one. Photoelectrons are efficiently focused into THGEM 1 and multiplied in two steps.

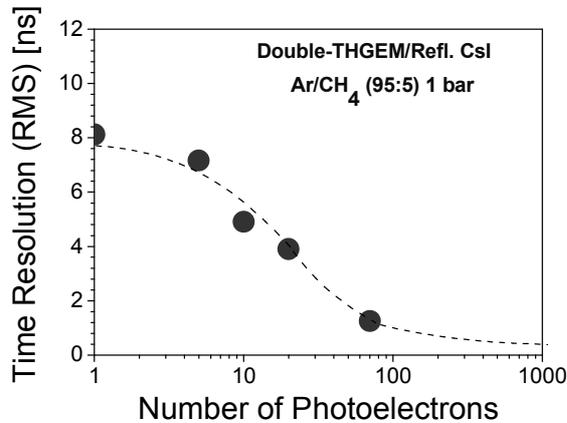

Figure 6. Time resolution (RMS) vs. number of photoelectrons recorded with a pulsed UV lamp in a reflective double-THGEM gaseous photomultiplier with CsI photocathode of figure 5.

The improved time-resolution with the number of photoelectrons results from measuring the "first-arriving photoelectron" (among those photo-produced at different locations on the photocathode's surface or arriving at different times due to diffusion [17]) and from improved signal-to-noise ratio.

Measurements with MIPs were done while converting radiation in a 3mm drift gap (Fig. 3). The THGEM's time-pulses were measured against scintillators, either with $^{106}$Ru beta-electrons or with cosmic rays. Both MIPs yielded similar time-resolution values (Fig. 7) of the order of 10ns (RMS). The tail in the time distribution is due to the statistical pulse-height distribution of single ionization-electron pulses, affecting the trigger electronics. Note that we obtained, with the same setup and electronics, very similar tail-shape and resolution (7-8ns RMS) with a standard triple-GEM detector. More details are given in [17].

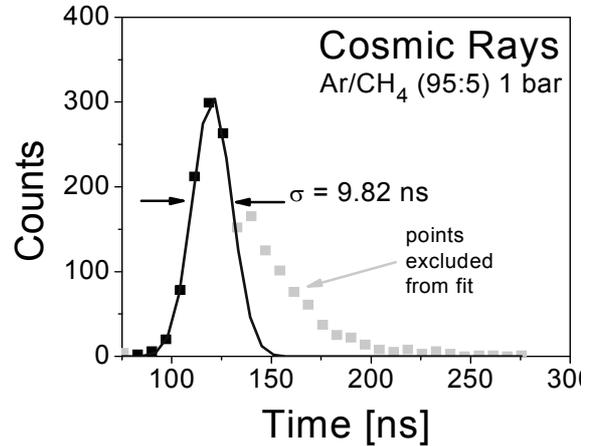

Figure 7. Time resolution of a double THGEM (of Fig. 3) measured with cosmic rays.

## 3. THGEM potential applications

The robustness, simplicity and properties of the THGEM and the possibility of industrial production capability of large-area detectors, pave ways towards a broad spectrum of potential applications. These could rely on THGEM's single-electron sensitivity, moderate (sub-mm) localization resolution, timing in the 10ns range, high-rate capability, low-temperature and broad pressure-range (mbar to few bar) operation.

Particle- and astroparticle-physics applications could encompass: tracking at moderate resolutions

(e.g. large-area muon- or cosmic-ray detectors), sampling-elements for calorimeters, large-volume TPCs for rare events, single-photon detectors for RICH, etc.

Large-area THGEM UV-photon detectors with reflective CsI photocathodes [11,13] (Fig. 5) would have some advantages over cascaded-GEM ones [23,24]. E.g. the better electron collection and transport between cascaded elements results in a lower gain required at each single-multiplier step or, alternatively, fewer cascaded elements for an equal total gain.. Like in reflective-GEM photomultipliers, a small reversed drift field above the photocathode reduces significantly the detector's sensitivity to charged particles (Fig. 8) [23,25].

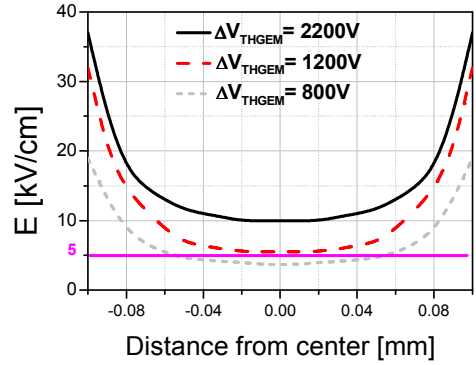

Figure 9. Electric field on photocathode surface versus distance from hole center, created by the hole dipole field in the reflective-photocathode photon detector of Fig. 5 THGEM parameters: d=0.3mm; a=0.7mm; t=0.4mm.

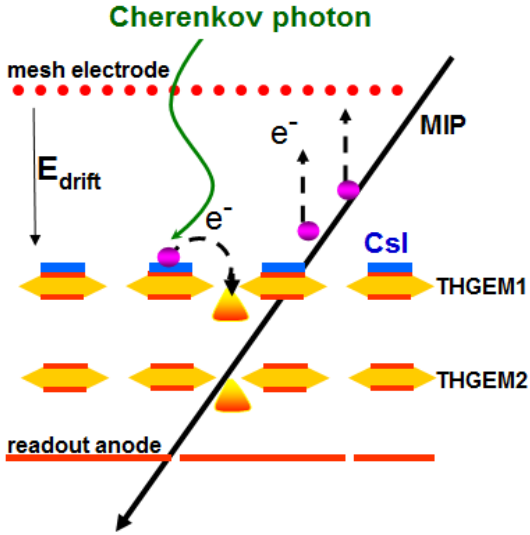

Figure 8. A double-THGEM gaseous photon detector for RICH. The mesh defines a small reversed drift-field above the photocathode, repelling a major fraction of ionization electrons; the photoelectron collection efficiency is very little affected.

The high electric fields at the THGEM's photocathode surface (between holes), reaching values of a few kV/cm (Fig. 9) assure good photoelectron extraction. The latter results in reasonably good effective-QE values also in noble-gas mixtures (Fig. 10) [26, 27].

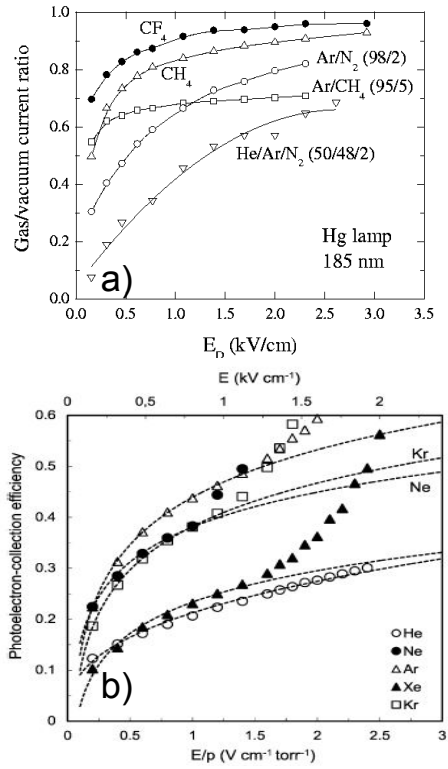

Figure 10. Measured photoelectron collection efficiency into different gases vs. electric field in: a) molecular gases and gas mixtures; b) noble gases. In the latter the dashed lines are model-calculated; the divergence measured in Ar, Kr, Ne and Xe for $E > 1.2kV/cm$ originates from avalanche-multiplication onset.

The operation of THGEMs in noble-gases has potential applications in gas-scintillation radiation detectors, two-phase (Fig. 11) and noble-liquid cryogenic detectors for rare events and for Gamma imaging. Cryogenic gaseous UV-photomultipliers with THGEM-coated CsI-photocathodes are under development for scintillation-light recording in some of these applications. This should pave ways towards development of novel large-mass detectors, having high sensitivity to rare events, with low-radioactivity background (e.g. compared to photomultiplier tubes), and low energy threshold – at moderate costs. Potential applications are in the fields of neutrino, double-beta decay and dark-matter physics.

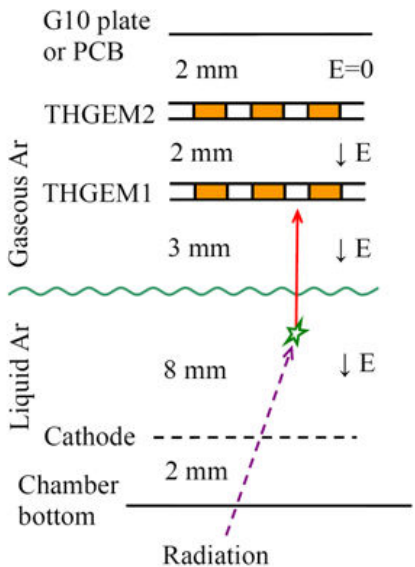

*Figure 11. Example of a double-THGEM cryogenic detector measuring radiation-charges extracted from noble-liquid. A photocathode deposited on the front face of THGEM1 (not shown) would also detect noble-liquid scintillation light [20].*

Other fields in which THGEM detectors are under R&D are: soft x-ray imaging (Fig. 3), thermal- and fast-neutron imaging with adequate converters coupled to the multiplier or deposited on its surface (similar to Fig. 5), etc.

Last but not least is the resistance to damaging discharges, e.g. in cases of single-electron (photoelectron) or low-ionization (MIPs) detection accompanied occasionally by heavily-ionizing background. Laboratory studies proved the THGEM to be robust and very resistant to sparks, compared to GEM.

A REsistive THGEM (RETHGEM) [27] was recently introduced, in an attempt to conceive a spark-immune multiplier. In the RETHGEM the Cu-clad (Fig. 1) is replaced by a resistive coating e.g. resistive Kapton, silk-screen printed surface etc. Like other detectors with resistive surfaces (e.g. RPCs) it has an improved resistance to discharges, but at the expense of lower counting-rate capability - of the order of 10-100Hz/mm$^2$. Gains $\geq 10^5$ were reached in different gases in double-RETHGEM coupled to a CsI photocathode [27].

**Acknowledgements**

This work was partly supported by the Israel Science Foundation, grant 402/05, the MINERVA Foundation grant 8566 and the Foundation for Science & Technology (FCT) – Portugal, project POCI/FP/81980/2007. We greatly acknowledge the assistance of V. Revivo and G. Genach, the PCB producers (Print Electronics, Israel, www.print-e.co.il). A. Breskin is the W.P. Reuther Professor of Research in The Peaceful Use of Atomic Energy